\documentclass[12pt]{article}

\usepackage{aaspp4}
\usepackage{epsfig}

\begin{document}

\begin{center}
\uppercase{{\sc Are Large X--Ray Clusters at Thermal Equilibrium ?}}
\end{center}

\author{\sc Jean-Pierre Chi{\` e}ze}
\affil{CEA, DSM/DAPNIA/Service d'Astrophysique, CE SACLAY, F-91191 Gif-sur-Yvette, France}

\author{\sc Jean-Michel Alimi}
\affil{Laboratoire d'Astrophysique Extragalactique et de Cosmologie, CNRS URA 173, Observatoire de Paris-Meudon, 92195 Meudon, France}

\author{\sc Romain Teyssier}
\affil{CEA, DSM/DAPNIA/Service d'Astrophysique, CE SACLAY, F-91191 Gif-sur-Yvette, France}

\begin{abstract}
We simulate the formation of a large X--ray cluster using a fully
3D hydrodynamical code coupled to a Particle--Mesh scheme which
models the dark matter  component. We focus on a possible
decoupling between electrons and ions temperatures. We then solve
the energy transfer equations between electrons, ions and neutrals
without assuming thermal equilibrium between the three gases ($T_e
\ne T_i \ne T_n$). We solve self-consistently the chemical
equations for an hydrogen/helium primordial plasma without assuming
ionization--recombination equilibrium. We find that the electron
temperature differs from the true  dynamical temperature by 20\% at
the Virial radius of our  simulated cluster. This could lead
marginally to an underestimate of the total mass in the outer
regions of large X-ray clusters.
\end{abstract}

\section{INTRODUCTION}

Large X-ray clusters are well  defined cosmological objects which
can provide useful  constraints on currently discussed structure
formation theories. It is believed that they are composed  mainly
of dark matter and X-ray emitting  gas.  The physical conditions of
the  hot gas are rather  extreme.  Common values for the electrons
density range from $10^{-1}$ cm$^{-3}$  in the core to $10^{-5}$
cm$^{-3}$ in  the outer regions.  The electrons temperature is
about $10$ keV, up to $15$ keV for A2163, the hottest cluster known
so far (Arnaud et al.  1992).

Recently, Markevitch et  al.  (1996) studied the electrons
temperature profile of A2163 using different X-ray experiments.
They measured $T_e \simeq 4$ keV  at a radius  corresponding
roughly to the Virial radius of  the cluster ($r_{200} \simeq 2.4$
Mpc h$^{-1}$). Using the well known hydrostatic equilibrium
equation  in order  to derive the total mass distribution,  they
conclude    that  the total mass distribution is significantly
steeper than the X-ray gas distribution. But, as  shown  by
Schindler \&  Muller (1993) and Evrard, Metzler \& Navarro (1996),
the hydrostatic equilibrium assumption is unlikely to be true in
the  low density regions. Moreover, at that  radius, the electron
density is supposed to  be roughly a few $10^{-5}$ cm$^{-3}$, and
the time-scale for  electrons to reach thermodynamical equilibrium
with  ions is then  about 4 Gyr,  comparable  to the merger
time-scale (Markevitch  et al. 1996).  Therefore,  one  can  ask
the following question: in the outer region of A2163 and more
generally in any large X-ray cluster, is $T_e = T_i$ ?   If $T_e
\ne  T_i$, this could result in additional errors  in  the mass
estimate   due to a departure  from thermodynamical equilibrium
between ions and electrons.

In Teyssier, Chi{\`e}ze  \& Alimi (1997),  as  a first approach to
this problem, we studied the collapse of a planar density
perturbation, usually called a Zel'dovich pancake, of comoving
wavelength $L=16$ Mpc h$^{-1}$.   This  rather  formal case enabled
us to test with high resolution 1D simulations our hydrodynamical
code which solves a set of collisional processes such as energy
exchange, non--equilibrium chemistry, shock heating and electronic
conduction. We showed that a large region of the pancake ($\simeq
1$ Mpc h$^{-1}$) does not recover thermodynamical equilibrium. Only
the central part of the pancake recovers within a  few percent $T_e
\simeq T_i$. The strongest departure from thermodynamical
equilibrium was found near the shock front, where $T_e$ is one
order of magnitude lower than $T_i$.

For a  real cluster, much  higher densities  and temperatures than
for pancakes are expected in the high temperature, X-ray emitting
gas. In this paper, we thus intend to make a quantitative study of
the thermodynamical history of the gas during the hierarchical
formation of a three-dimensional (3D) X-ray cluster embedded in a
standard CDM cosmogony.  We solve  precisely the energy transfer
equations between electrons, ions and  neutrals, in order to
confirm or infirm the thermodynamical equilibrium assumption.

To test the thermodynamical evolution of a realistic X-ray cluster,
we have developed a  3D  hydrodynamical code, called HYDREL
(``HYDrodynamique Euler Lagrange"), coupled to  a  PM scheme which
describes the  dark  matter component. In section  2 we describe
the physical processes involved in the primordial collisional
plasma, the numerical methods we use to simulate the formation of
X-ray clusters. and the initial conditions  we choose. In  section
3 we  present our results,  showing that significant departure from
thermodynamical equilibrium can be obtained at a radius $r
\simeq r_{200}$\footnote{$r_{200}$ is the radius of the sphere
centered on the cluster center and containing a mean over--density
$\bar{\delta} = 200$}. We discuss the general properties of the
simulated cluster, which has several average characteristics
similar to those of the Coma cluster. Finally, in section 4, we
discuss  possible observational consequences of our work, such as
an underestimation of the total mass in the outer regions of large
X-ray clusters.

\section{PHYSICAL AND NUMERICAL METHODS}

In this paper, we intend to give a self-consistent description of
the thermodynamical evolution of a cluster of galaxy embedded in an
Einstein-de Sitter universe. We use the so--called ``standard CDM
cosmogony" with $\Omega=1$, $\Omega_B=0.1$ and $h=0.5$. This
scenario is one of the most typical example of the hierarchical
clustering picture. We use the Bardeen et al. (1986) power spectrum
to generate our initial Gaussian random field. In this section, we
briefly recall the physical processes that we study here. We also
present our numerical algorithm, namely the 3D hydrodynamical code
HYDREL and we finally discuss the numerical parameters that we use
for the specific realization presented here.

\subsection{Physical Processes}

The aim of this paper is to study the effect of several processes
which we believe to be relevant in the primordial collisional
plasma. We therefore focus on collisional processes and distinguish
three thermodynamical species, namely electrons, ions and neutrals.
Each specy is supposed to be individually at the local
thermodynamical equilibrium (LTE), but we allow $T_e\ne T_i\ne
T_n$, where the subscripts design electrons, ions and neutrals
respectively. The LTE hypothesis is valid since the
``isotropization time-scale" $t_{iso}$, which drives the
distribution function of a given specy to a Maxwellian, is very
small, even in the rather extreme conditions found in X-ray
clusters. Indeed, from $n_e\simeq n_i=10^{-5}$ cm$^{-3}$ and
$T_e\simeq T_i$ = $10^7$ K, we can deduce an estimation of $t_{iso}
\simeq 10^{5}$ yrs for electrons and $t_{iso} \simeq 10^{6}$ yrs
for ions (Spitzer 1962). On the other hand, we outlined in the
introduction that the ``equipartition time-scale" between electrons
and  ions was quite long in the outer regions ($t_{ei} \ge 10^{9}$
yrs). We  must therefore solve the energy transfer equations
between the three thermodynamical species. Note that in the case of
X-ray clusters, the neutral component is of weak relevance, since
the medium is fully ionized. However, our code was designed for a
more general use, and therefore we follow self--consistently all
chemical species.

For electrons and protons, the internal energy transfer per unit
volume and per unit time is due to Coulomb collisions and writes
(Spitzer 1962)

\begin{equation}
\frac{\delta {\cal Q} }{\hbox{D}t} =  -n_{e}n_{p}k
\left(T_{i}-T_{e}\right) \left(
\frac{4(2\pi)^{1/2}e^{4}m_{e}^{1/2}\ln{\Lambda _{ep}} }
{m_{p}\left( kT_{ei} \right) ^{3/2} } \right)
\label{eqei}
\end{equation}

\noindent
where  $T_{ei}$ is ``the reduced  temperature" of the two
interacting particles and $\Lambda_{ep}$ is  the Coulomb logarithm.
This formula can be applied  to other ions   with of course  a
modification due to their  different atomic masses and   charges
(Spitzer 1962).  We  also compute the energy exchange rate between
electrons and neutrals, using the classical ``hard reflecting
sphere" cross-section  $\sigma_{en} \simeq 10^{-15}$ cm$^2$ (Draine
\& Katz  1986).  The  energy exchange between ions and neutrals
which is due to the resonant charge transfer interaction, is
derived  by using the momentum transfer cross--section of HI--HII
presented in Hunter \& Kuryan (1977) and Draine (1980).

To compute accurate  energy   exchange rates,  we solve the
chemical equations without  assuming ionization--recombination
equilibrium.  We consider    only  6  chemical  reactions,  which
are  ionization and recombination  for HI--HII, for  HeI--HeII and
for HeII--HeIII.  This non--equilibrium approach is important
behind   shock fronts, where  a description using the Saha equation
would overestimate the ionization fraction, and would lead to wrong
abundances and wrong energy exchange rates.  We use the chemical
reaction rates presented in Cen (1992).

In rich galaxy clusters, the gas temperature ranges from 1  to 10
keV. Line cooling is therefore negligible. We only take into
account Brehmstrahlung (Mewe et al. 1987) and Compton cooling by
the Cosmic Background Radiation (Peebles 1993). These cooling
processes are likely to lower slightly the electrons temperature,
and therefore enhance the departure from thermodynamical
equilibrium.

In Teyssier et al. (1997), we also considered the influence of
electronic conduction, assuming no transverse magnetic field. We
showed that conduction is effective only in the very low density,
outer regions, where a thermal precursor preheats the gas ahead of
the shock front (Zel'dovich \& Raizer 1966). The downstream flow
properties were qualitatively similar to the non conductive case,
altough the temperature decoupling between ions and electrons was
slightly lowered. This led us to conclude that, even in the
assumption of no magnetic field, electronic conduction might have
no direct observational consequences. On the other hand, it has
been shown that a small magnetic field ($\simeq$ 1 $\mu G$) does
exist in the intracluster medium (Kim et al. 1990). Electronic
conduction should then be efficently suppressed. We therefore do
not consider it in the present paper.

The presence of a magnetic field does not affect classical
collisional energy exchange, since it is a purely local process.
However, a magnetic field has also the well-known effect of
introducing various plasma instabilities within the shock
structure. These instabilities  are believed to be responsible of a
rapid, anomalous heating of electrons. Cargill \& Papadopoulos
(1988) proposed a mechanism for this strong, collisionless heating,
based on the Buneman and ion acoustic instabilities. This mechanism
justifies why electrons are efficiently heated in young supernovae
remnants, an observationnal fact which was unexplained by the
classical collisional theory. However, Cargill \& Papadopoulos
(1988) showed that only 12\% of the upstream kinetic energy can be
converted into electrons thermal energy. Consequently, after the
collisionless shock front, complete ions-electrons equipartition
still relies on classical collisional processes. In the
calculations we present in this paper, altough we neglect the
various plasmas instabilities discussed here, we obtain rather low
temperature differences, namely $T_e \ge T_i/5$. This justifies the
use of classical collisions theory only.

Keeping in mind the physical assumption we just made, let us now
summarize the thermodynamical history of the intracluster gas. Ions
and neutrals are shock heated through mergers or accretion shock
waves. Electrons are less efficiently heated by shocks, as can be
shown by hand using the Rankine--Hugoniot discontinuities
relations. As a matter of fact, these relations state that a gas
with mean molecular weight $\mu$ is heated by a shock front with
velocity $D$ up to a post--shock temperature given by

\begin{equation}
kT = \frac{3}{16}\mu D^2
\end{equation}

\noindent
in the limit of very high Mach number. Just after the compression
front, electrons temperature is therefore much lower than ions
temperature $T_e \simeq  (m_e/m_p)T_i$, where the pre-factor on the
r.h.s is of the order of $10^{-3}$. The plasma finally reaches
thermodynamical equilibrium ($T_e \simeq T_i$) after a few
ions--electrons energy exchange time--scales $t_{ei}$, given  by
equation (\ref{eqei})

\begin{equation}
t_{ei} \simeq 503 \frac{T_e^{3/2}}{n_e} \hbox{ sec}
\label{tei}
\end{equation}

The length of this so--called ``equipartition wave" where a
significant departure from thermodynamical equilibrium is expected,
can be estimated using $L_{ei} \simeq (1/4) t_{ei}D$. Under the
rather extreme physical conditions encountered in large X-ray
clusters, $n_e \simeq 10^{-5}$ and $D \simeq 1000$ km s$^{-1}$,
this ``equipartition mean free length" is very extended,
$L_{ei}\simeq 1.4$ Mpc h$^{-1}$. Moreover, one clearly sees from
equation (\ref{tei}), that the higher is the gas temperature, and
the lower is the gas density, the larger is this equipartition
region. In Teyssier et al. (1997), we calculated more precisely the
size of this region for a pancake of initial comoving wavelength $L
= 16$ Mpc h$^{-1}$, and we found $L_{ei} \simeq 1$ Mpc h$^{-1}$
with a total shocked region of $2$ Mpc h$^{-1}$.

In a fully 3D environment, shock waves  interact in a very
complicated pattern. Hierarchical merging  means here that small,
low temperature sub--structures merge together, leading to strong
shocks propagating in a low density environment, especially in the
outer part of  X-ray clusters. Consequently, we need a 3D
hydrodynamical code which self--consistently solves the gas
dynamics equations with the different collisional processes
previously mentioned.

\subsection{Numerical Schemes}

The choice of our numerical method is dictated by the specific
regions of clusters  we are interested in. These regions are likely
to be far from thermodynamical and chemical equilibrium. First, the
low value of the gas density results in rather slow collisional
time--scales. Second, these regions  are not relaxed, with  high
bulk velocities and strong shock waves, making non--equilibrium
phenomena dominant. We therefore use an Eulerian hydrodynamical
code (Kang  et al. 1994) to simulate the formation of a large
cluster of galaxies.

Our code, called HYDREL, has been presented  in great details for
its 1D version in Teyssier et al. (1997). We briefly recall here
its main characteristics, as well as the specific 3D features.
HYDREL is based on an operator splitting algorithm, and  solves the
different thermo- and hydrodynamical equations in 3 consecutive
steps. The first step, called the gravity step, solves the Poisson
equation. It calculates the gravitational potential deduced from
the gas  and dark matter density fields. Dark matter particles are
displaced during this step with a classical Particle--Mesh  (PM)
scheme (Hockney \& Eastwood 1981), developed by Alimi and Scholl
(1993) first on Connection Machine, then implemented on Cray--YMP.
The equation of motion in this PM code are solved in comoving
coordinates and the Green function takes into account aliasing
effects and minimizes force's anisotropies. The time integrator of
the PM has however been modified, it is now based on a
predictor--corrector scheme. This allows both great accuracy and
variable time--stepping which is impossible with the classical
Leap--Frog scheme. The second step is the adiabatic hydrodynamical
step. It solves the hydrodynamical equations using directional
splitting and a staggered  mesh. Shock waves are treated using the
pseudo--viscosity method (Von Neumann \& Richtmyer 1950). We use
for that purpose a viscous tensor, and not a viscous pressure. This
tensorial formulation (Tscharnuter \& Winkler 1979;  Mihalas \&
Mihalas 1984; Stone  \& Norman 1992; Chi{\`e}ze et al. 1997) is of
great importance for cosmological flows. We therefore recall now
the main features of our tensorial pseudo--viscosity in 3D.

This approach relies on the assumption that dissipation in shock
waves is correctly  described by the Navier--Stokes equations. For
each direction $i=1,2,3$, we use a  diagonal  stress tensor, whose
coordinates $\sigma _i$, are proportional to the diagonal terms of
the shear tensor

\begin{equation}
\sigma_i = \frac{P}{c_s}\Delta x
\left(\frac{\partial u_i}{\partial x_i} - \frac{1}{3}\nabla \cdot u
\right)
\end{equation}

\noindent
where $\nabla \cdot u$ is the  divergence of the velocity field,
$c_s$ is the local sound speed and $P$  the thermal pressure.
Non--diagonal terms in  the usual Navier--Stokes stress tensor  are
dropped here, in order  to  avoid spurious  turbulent   effects.
Indeed,  in  the last formula, $\Delta x$ is the mesh size.  For  a
real viscous fluid, this term  has  to  be replaced   by the mean
free  path $l$ of   the gas particles, which is orders of magnitude
lower than the cell size.  The term ``pseudo--viscosity"  rely on
the artificial enhancement of the mean free path ($l \rightarrow
\Delta x$) due to the finite resolution of the grid.  The  viscous
stress exerts  on each fluid element a  net force given by

\begin{equation}
F_i = - \frac{\partial}{\partial x_i} \sigma _i
\end{equation}

\noindent
Note that the three components  of the pseudo--viscous force differ
in general.  A viscous pressure, in the contrary,  would have been
always isotropic.  Note also that in case of  an homologous flow
($\sigma_i = 0$), there is no dissipation in the flow (Mihalas \&
Mihalas 1984).  A viscous pressure would not have  been able to
satisfy this fundamental physical requirement.  Because shock waves
result  in a contraction of the fluid  elements they cross,   the
pseudo--viscous force  acts only when the following criterion is
full-filled

\begin{equation}
\frac{\partial u_i}{\partial x_i} < \frac{1}{3}\nabla \cdot u < 0
\end{equation}

The  hydrodynamical step we describe  now is divided  in two
sub-steps for  each direction: first the gas  dynamics equations
are solved in a Lagrangian  way, then  we  re-map  the  new  flow
variables from   the perturbed Lagrangian grid to  the fixed
Eulerian one.   The Lagrangian step   uses a predictor--corrector
time  solver, which ensures second order accuracy in time.   The
Eulerian step uses  the Van  Leer (1977) advection scheme, which
ensures second order accuracy in space.

The  third  step  is  called  the  dissipative  step.  It   solves
all collisional  processes,  namely  chemical  reactions,  energy
exchange between the three thermodynamical processes and cooling.
Because these processes are  driven by rather stiff  equations, we
use for each cell an individual time--stepping.  This  allows us to
compute new energies and  abundances with  high  accuracy without
slowing  down the  whole simulation. Chemical reactions and the
other collisional processes are strongly coupled equations.  To
ensure stability, we solve the system of chemical and
thermodynamical equations using the ``fully  implicit method". The
large hydrodynamical  time--step is  controlled by the Courant
condition, and   the small sub--time--steps are  controlled by the
relative variations of the  chemical and thermodynamical variables
in each cell.  In the  case of strong  cooling ($t_{cool} <
t_{dyn}$), the pressure can be  dramatically underestimated.  We
therefore impose that  during the dissipative step,   the total
pressure does not  vary more than 10\%. This method  works well in
general,  and allows a very good accuracy  in   the chemical
calculations.    However, for  large clusters of galaxies, which
are the purpose of  this paper, cooling is not  efficient, due to
the  high temperature of  the intra-cluster gas ($T > 1$ keV) and
the relatively  low--densities that we obtain in our simulation.

\subsection{Initial Conditions}

We now present the  initial conditions that  we use to simulate a
rich cluster  of galaxies embedded in a  CDM cosmogony.  We  use a
comoving box size of $25$ Mpc h$^{-1}$,  with periodic boundary
conditions.  It would  have been better to use  a larger box,  but
we have  to make a compromise between large scale power  and
spatial resolution.  We made a choice similar to Anninos \& Norman
(1996),  and thus we can compare directly our  results to  their
calculations.   We use   $128^3$ grid points, and the  same number
of dark  matter particles.  The numerical force is 50 \% of  the
true gravitational force  at a scale of roughly 1.5 cells.  This
gives an effective resolution  of $300$ kpc h$^{-1}$. This
corresponds also to the hydrodynamical   resolution, as shown by
extensive tests (Teyssier et al.  1997).

We start  our   simulation at a  redshift  $z_i  = 50$.    The
initial abundances  are   taken from  Peebles  (1993)   for our
chosen value $\Omega_B = 0.1$.  The gas temperature is initially
uniform and we use the relation $T(z_i) = (1+z_i)^2  1.37 \times
10^{-2}$ K, which states that the gas temperature is strongly
coupled to the CBR up to $z=200$, and then evolves adiabatically up
to $z_i$.  Dark matter particles are initially uniformly
distributed on the grid,  and then displaced using the Zel'dovich
approximation.   The  initial baryons density field  is supposed to
be equal to the initial Gaussian random density field.  We impose a
$3\sigma$ peak of length scale $4$ Mpc h$^{-1}$ at the center of
the  box, using the  Hoffman--Ribak (1991) algorithm.  We reach the
final epoch (defined as the epoch when the linear r.m.s. is equal
to 1 at 8 Mpc h$^{-1}$)  with approximately 350 time--steps,
controlled by the Courant condition.   The  energy conservation, as
defined   by Cen (1992), was less than 0.8 \% during the whole run.

\section{RESULTS}

The cluster we obtained at $z = 0$ have average characteristics
(mass, size and temperature) similar to those observed for the Coma
cluster. We define the cluster center as the cell of maximum X--ray
emissivity. We then define the Virial radius $r_{200}$ of the
cluster as the radius of the sphere centered on the cluster center
and containing a mean over--density $\bar{\delta} = 200$. We find
in this way, $r_{200}=1.6$ Mpc h$^{-1}$. The total mass embedded in
this radius is $M_{200} = 9.8 \times 10^{14}$ M$_{\odot}$ h$^{-1}$.

\begin{figure}[httb]
\hbox{\psfig{file=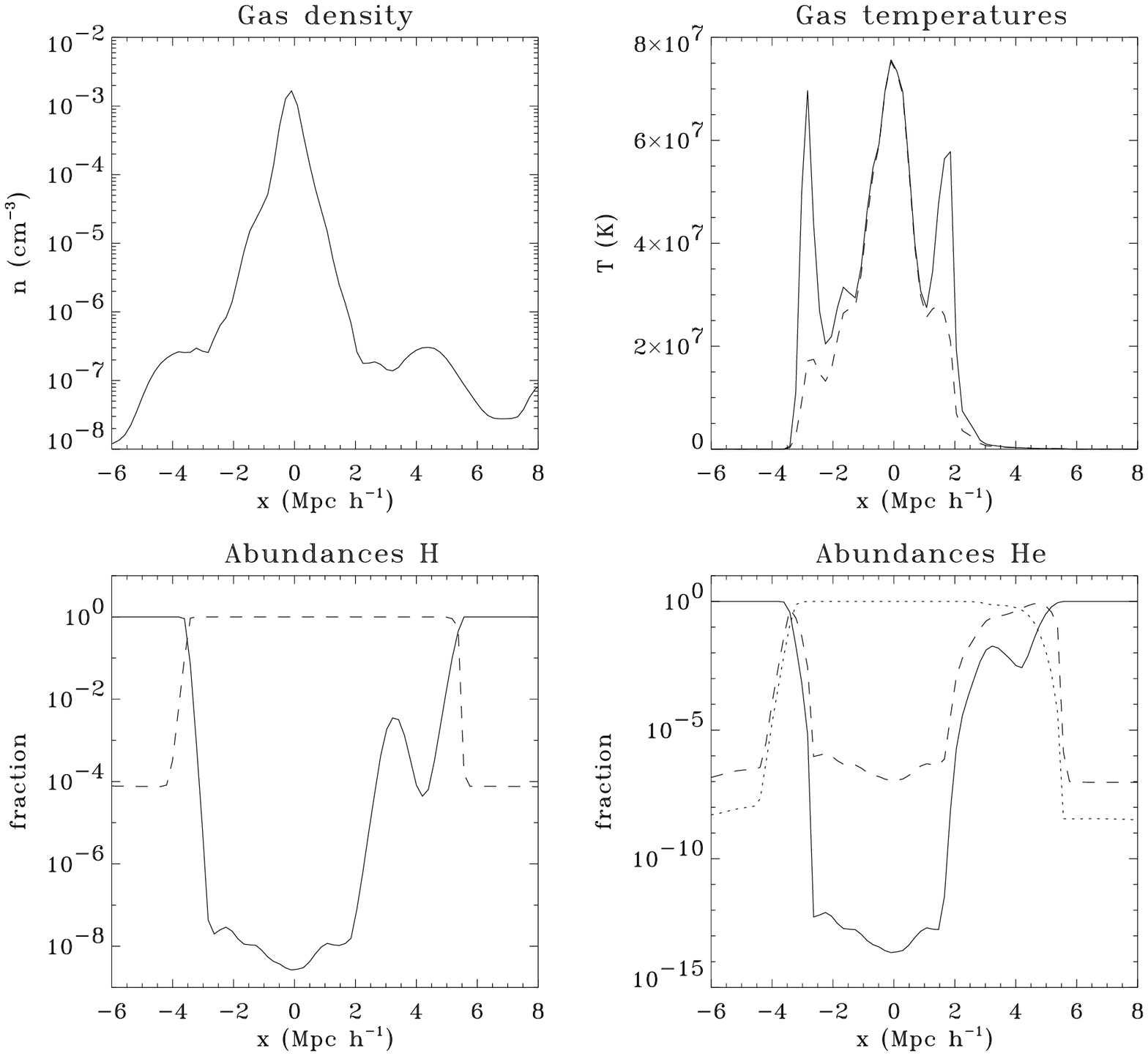,height=13cm}}
\caption{ Gas density in atomic mass per cm$^3$,  temperatures: ions
(solid  line)  and electrons  (dashed   line), abundances of
Hydrogen species: fraction of HI (solid line) and HII (dashed
line), abundances of Helium species: fraction of   HeI (solid
line), HeII (dashed  line) and HeIII  (dotted line). The different
quantities  are taken  along a line of sight which crosses the
cluster center.}
\label{cutline}
\end{figure}

We plot in figure (\ref{cutline}) the gas density, temperatures and
ionization fraction along a line of sight which crosses the center
of the cluster. We can define here three characteristic regions in
the vicinity of the cluster:
\begin{itemize}
\item the  cluster itself,  which   has recovered thermodynamical  and
  chemical equilibrium,
\item a  large  non--equilibrium  region,  where  ions  and  electrons
  temperatures   differ significantly,
\item and finally the cold, unshocked inter--cluster medium.
\end{itemize}

Note that the accretion shock is relatively steep, and that the gas
is very quickly ionized. We did not consider here any ionizing
background, altough it is strongly suggested by the Gunn--Peterson
effect, but ionization by  shock waves turned out to be efficient
enough, as soon as the intracluster medium is concerned. The
equipartition front, where $T_e$ gradually reaches $T_i$, has a
thickness of approximately 1.5 Mpc h$^{-1}$. This non--equilibrium
region stands mainly outside the Virial radius of the cluster. The
shock front, which marks the beginning of the non--equilibrium
region, is located at roughly $2 r_{200}$ of the cluster center.
The temperature decouling between ions and electrons is maximum
just after the shock front, but always greater than $T_e \simeq
T_i/3$ (except ``inside" the shock front). This justifies a
posteriori our assumption that plasma instabilities could be
neglected here (see Section 2.1).

\begin{figure}[httb]
\hbox{\psfig{file=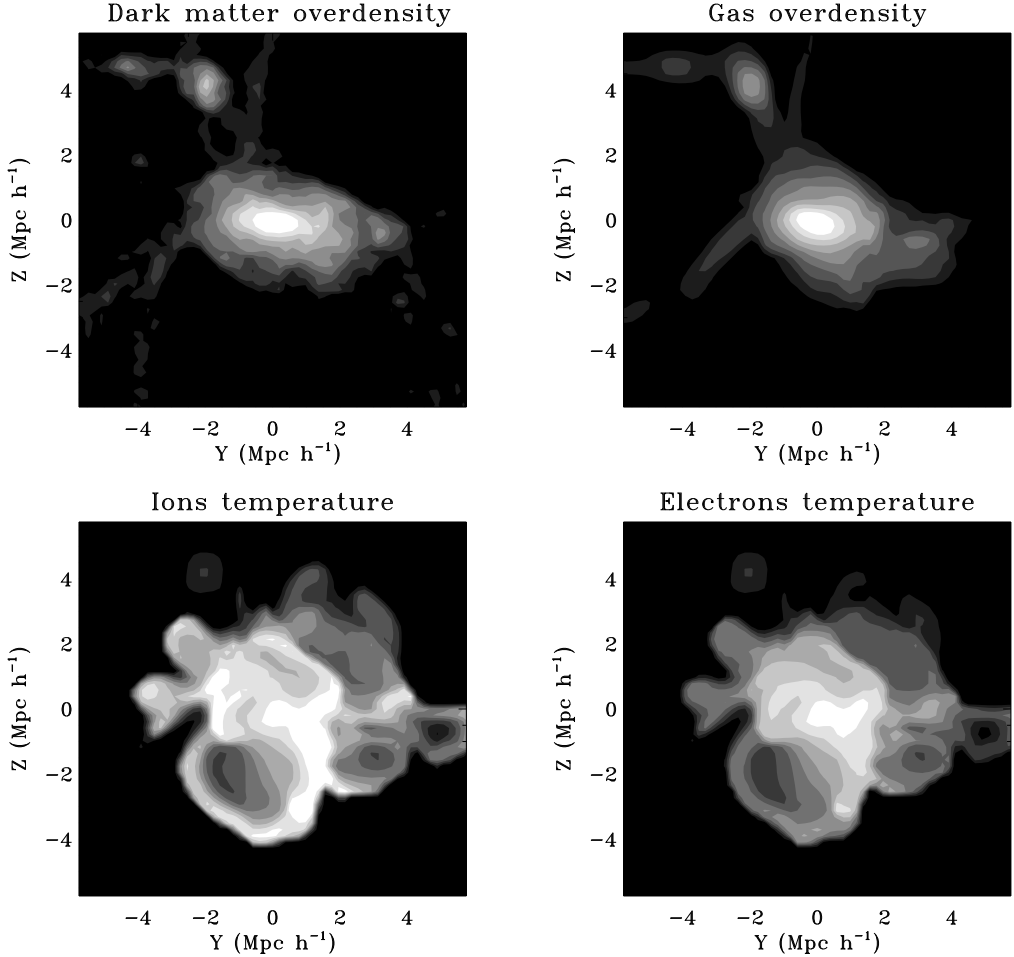,height=13cm}}
\caption{  Upper panels: gray scale images  of the gas and dark matter
density  contrast  with 10 levels,   defined by a constant
logarithmic spacing between $\delta = 1$ and $\delta  = 10^{3.5}$.
Lower pannels: gray  scale images  of the  electrons  and ions
temperatures with  10 levels, defined by a constant logarithmic
spacing between $T = 10^6$ K and $T = 10^8$ K. The slice is
one-cell width and crosses the cluster center.}
\label{contours}
\end{figure}

In figure (\ref{contours}), we plot gray scale images of the dark
matter density contrast, the gas density  contrast and the
electrons and ions temperatures. Note that the filamentary
structures clearly converge  towards the cluster. Gas and dark
matter isocontours have similar elliptical shapes, with axis ratio
2:1. They are both relatively smooth. In the contrary, the
temperatures isocontours show very complicated patterns, with
several shock waves propagating in different directions. Note that
the electrons temperature appears much smoother than the ions
temperature. The hottest regions are not located in the center  of
the cluster, but in the outer regions where strong shock heating
occurs. This explains why the strongest temperature decoupling is
mainly located in the low density, outer regions of the cluster.

\begin{figure}[httb]
\hbox{\psfig{file=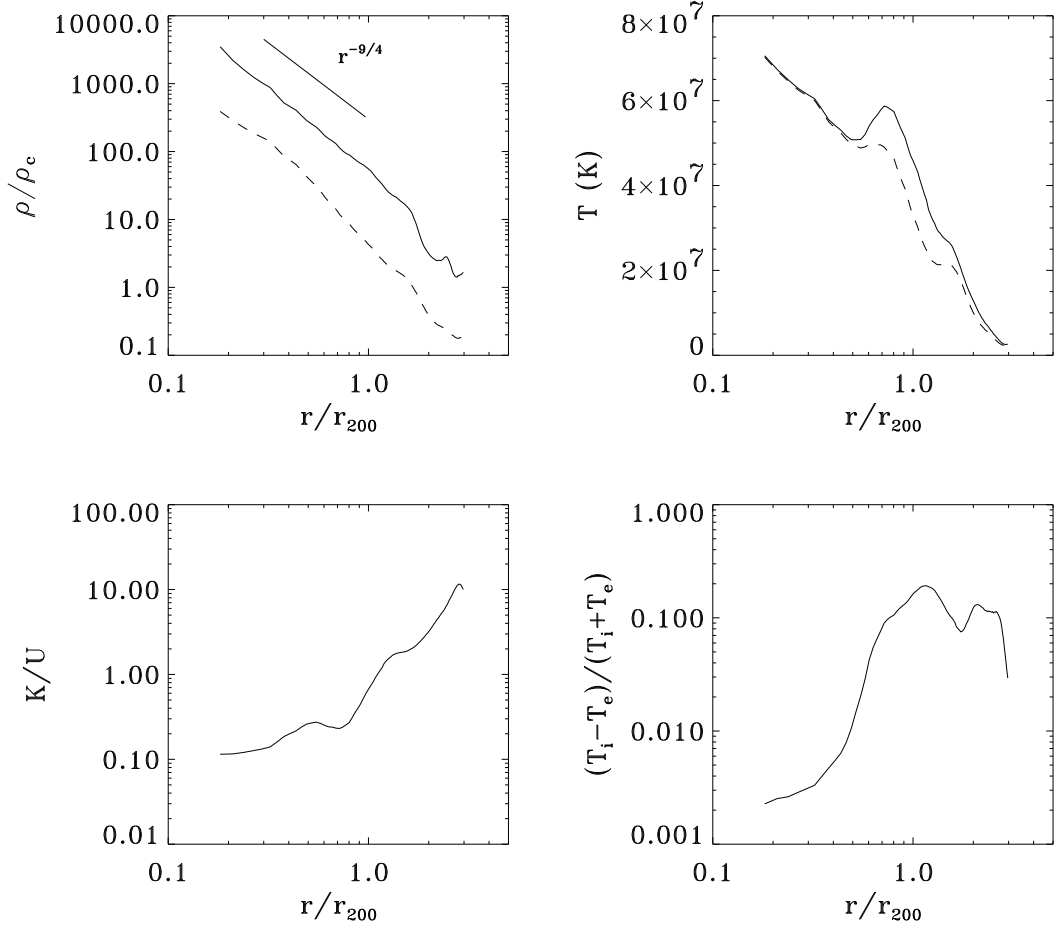,height=13cm}}
\caption{ Upper left pannel: spherically  averaged density profile for
dark matter  (solid line) and  gas   (dashed line) versus  the
radial distance to the cluster center,  in units of $r_{200}$. The
$r^{-9/4}$ power  law  is also shown   as a straight   line.  Upper
right pannel: spherically averaged temperature for electrons
(dashed line) and ions (solid  line).  Lower  left pannel:  ratio
between the bulk  kinetic energy and the internal energy of  the
gas.  Lower right pannel: ratio $(T_i-T_e)/(T_i+T_e)$ that measures
the departure from thermodynamical equilibrium (see text).}
\label{profil}
\end{figure}

We plot in figure (\ref{profil}) the spherically averaged density
and temperatures profiles. The radius is expressed in units of
$r_{200}$. The most inner point corresponds to our resolution limit
in the computation of the gravitational force. Note that gas and
dark matter density profiles are both very similar. We show no
evidence of core radii in any mass distribution. Moreover, both
density profiles are  well fitted by a  power law $\rho \propto
r^{-9/4}$. This is in good  agreement with Anninos \& Norman
(1996), who studied the influence of numerical effects on gas and
dark matter density profiles, using higher resolution simulations.
The electrons and ions temperatures profiles show again clearly
that a large non--equilibrium region extends from $r_{200}/2$  up
to $2r_{200}$. In order to quantify the error that one observer
does between the X-ray temperature $T_e$ and the true dynamical
temperature $T \simeq (T_e+T_i)/2$, we plot the ratio $(T-T_e)/T$
as a function of radius. The maximum departure from thermodynamical
equilibrium is located at $r_{200}$ and is about 20\%. We also
calculate the ratio between the bulk kinetic energy and the
internal energy of the gas. This ratio is equal to unity at
$r_{200}$, showing that hydrostatic equilibrium is also not
recovered in this region. Therefore, the hydrostatic equilibrium
assumption and the thermodynamical equilibrium assumption are both
valid in the central region of the cluster ($r<r_{200}/2$), but are
both violated in the outer regions of the cluster ($r>r_{200}/2$).

\section{CONCLUSION}

In this paper, we studied the formation of a rich  X--ray cluster.
We found that the total mass embedded in the Virial radius
($r_{200}=1.6$ Mpc  h$^{-1}$) was equal  to $10^{15}$ M$_{\odot}$
h$^{-1}$.  We found also that  the density profiles  of gas and
dark matter are both well fitted    by a  $r^{-9/4}$   power   law.
We therefore  have similar conclusions    than Anninos \&    Norman
(1996), using similar initial conditions.  We  studied more
specifically the thermodynamical history of the  intra--cluster
gas.   We found  that a significant  decoupling between electrons
and ions temperatures occurs between $r_{200}/2$ and the shock
front, located roughly at $2r_{200}$.  The maximum departure is
found at  $r_{200}$  and   reaches  20\%.  Therefore,   the  usual
assumption  of thermodynamical equilibrium  between ions and
electrons breaks   down in this region.  We   also  checked that
the hydrostatic equilibrium   assumption was not  valid in  the
outer  regions of the simulated cluster.    These  two    errors
can  both  lead    to   an underestimates  of  the gravitational
mass  in the  outer  regions of X--ray clusters ($r \ge
r_{200}/2$).  These results could be carefully extrapolated to the
case of A2163.   As mentioned in the introduction, Markevitch et
al.   (1996) measured at  the  Virial radius of  A2163 a
temperature  of 4 keV, which is  roughly equal to   twice the value
we found here   for our  simulated    cluster.  This  could  lead
to   a thermodynamical  decoupling of 50 \%,  which means that the
(observed) electrons  temperature  underestimates by a    factor of
two  the true dynamical temperature.  Therefore, for  A2163,  the
error in the  mass estimate due to a departure  from
thermodynamical equilibrium could be as large as a factor of two.
Further studies are however required to confirm these conclusions,
using other types of initial conditions.

We would like to thank R.  van den Weygaert for providing us the
code which generates constrained  realizations of  Gaussian random
fields  (van den Weygaert \& Bertschinger 1995).

\pagebreak

\end{document}